\documentclass[12pt]{article}
\usepackage{color,epsfig}
\usepackage{graphicx}

\def\be{\begin{equation}}
\def\fe{\end{equation}}

\begin{document}

\title{ {\small Contribution to 
{\bf Micro and Macro Structures of Spacetime}, Peyresq, June, 2004.}
\vskip 0.2 cm
{ Frozen rigging model of the energy dominated universe.}}

\author { {\bf Brandon Carter. }\\
 \hskip 0.2 cm\\   LuTh, Observatoire Paris-Meudon, 92195 France.}

\date{\it February, 2005}

\maketitle

\noindent
{\bf Abstract.\ } Composite rigging systems, involving membranes that meet
on strings that meet on monopoles, arise naturally by the Kibble mechanism 
as topological defects in field theories involving spontaneous symmetry 
breaking. Such systems will tend to freeze out into  static lattice type 
configurations with energy contribution ultimately be provided by the 
membranes. It has been suggested by Bucher and Spergel that on scales 
large compared with the relevant (interstellar separation) distance 
characterising the relevant mesh length, such a system may behave as a 
rigidity - stabilised solid, having an approximately isotropic stress 
energy tensor with negative pressure, as given by a polytropic index 
$\gamma=w+1=1/3$. It has recently been shown that such a system can be 
rigid enough to be stable if the number of membranes meeting at a junction 
is even (though not if it is odd). Using as examples an approximately
O(3) symmetric scalar field model that can provide an ``8 color'' (body 
centered) cubic lattice, and an approximate U(1)$\times$ U(1) model 
offering a disordered ``5 color'' lattice,  it is argued that such a 
mechanism can account naturally for the observed dark energy dominance of 
the universe, without ad hoc assumptions, other than that the relevant 
symmetry breaking phase transition should have occurred somewhere about 
the Kev energy range.

\noindent

\section{Introduction}

Kibble's idea that p-branes -- meaning structures with support confined 
to a thin neighbourhood of a (p+1) dimensional worldsheet -- may form 
naturally, as vacuum analogues (in field theories involving spontaneous 
symmetry breaking) of the kind of topological defect that is familiar in 
condensed matter physics~\cite{Kleman95}, has attracted widespread 
interest in recent years~\cite{Kibble95}, particularly in the context 
of cosmology, for which the consequences may be important.  The most 
commonly discussed possibility is that of cosmic string formation during 
Grand Unification, but there are many other eventualities whereby other 
later lower energy phase transitions could have given rise to vacuum 
defects of many other kinds. As well as scenarios giving rise to isolated 
monopoles or simple (domain wall type) membranes, there is a rich variety 
of more complex possibilities, such as strings with monopole endpoints 
or intersections, and membranes with string boundaries or junctions. 
Furthermore the strings and membranes involved need not necessarily be 
of simple Dirac-Goto-Nambu kind -- with action just proportional to their 
worldsheet measure giving a tension $T$ say that is constant and equal 
(in relativistic units) to the mass density -- but might also be of more 
complicated kind with variable (and in the membrane case anisotropic) 
tension due to internal currents of the kind first suggested by Witten 
in the case of cosmic strings, whose loops can thereby be stablised
as vortons~\cite{DavShell89}.
  
Convenient mathematical machinery is now available~\cite{Carter95} for 
treating the dynamical evolution, at a local level, of a system of such 
p-branes provided they behave collectively as a ``regular rigging'' 
structure (using terminology suggested by the nautical analogy with a 
system of ropes and sails) in the sense that each p-brane worldsheet 
interacts only by direct contact with p-1 branes forming its boundary 
or p+1 branes of which it is itself  a boundary. A prototypical example
of the application of this machinery is illustrated~\cite{CBA02} by the 
case of drum vortons. However, as in the familiar case of point particles 
(i.e. zero-branes) so also more generally, in the thin brane limit, 
technical problems will turn up due to divergences if it is necessary to 
allow for interactions between branes with dimension differing by 2 or 
more. In particular there will be divergence problems whenever it is 
necessary to allow for long range self interaction via the intermediary of
electomagnetic and gravitational fields, whose support is the underlying 
4 dimensinal spacetime manifold, which, considered as a continuous medium,
is effectively a 3-brane. 

Such an underlying 3-brane can interact in a well behaved 
manner (as described by Darmois-Israel type~\cite{BatCart01} junction 
conditions) with an ordinary membrane, which counts as a 2-brane, but not 
with strings and point particles. Provided that (as will usually be the 
case for the topological defects in question) the coupling of the long 
range self interactions is weak enough for a linearised approximation 
to suffice, the effect of the divergences can be allowed for at a 
local level in point particles, and to some extent in 
strings~\cite{CarBat98,CarBatUz03} by appropriate regularisation and 
renormalisation procedures. However a satisfactory treatment of the finite 
adjustment terms needed to allow for the effect of radiation back-reaction 
in strings remains elusive -- even for scalar axion radiation and linear 
electromagnetic radiation, and a fortiori for the gravitational radiation 
whose effects are important~\cite{Shellard95} for the long term evolution 
of a heavy string distribution.
 
Issues concerning heavy strings have in recent years come to seem less
important than previously, since trends in observational cosmology now 
disfavour the idea that GUT strings played a decisive role in galaxy 
formation. Attention has been focussed instead on a new issue which is 
that of the evidence for large scale cosmic acceleration implying the 
presence of a predominant ``dark energy'' constituent with a negative 
mean pressure $P$. The required value is so large -- compared with the 
apparent mass density $\rho$, which itself includes a substantial dark 
matter contribution -- that doubt has even been cast~\cite{CarrHofTrod03}
on such a generally accepted assumption as the ``dominant energy 
condition'' of Hawking and Ellis,  on which the demonstration of their 
(classical) vacuum conservation theorem is based~\cite{Carter03}. Although 
this need not apply by quantum fluctuations on a microscopic scale, it is 
indeed hard to see how a catastrophic instability of the vacuum could be 
avoided in a theory violating this condition, to the effect that at a
macroscopic average level,
\be \rho \geq \vert P\vert \, . \label{1}\fe

The present situation~\cite{CMMS04} is that the available evidence does 
not seem to impose quite such a radically revolutionary step as  
abandonning the usual supposition (\ref{1}), nor even to
impose the marginal limit $ \rho =\vert P\vert =-P$ requiring invocation
of a cosmological constant (whose magnitude would be hard to account for 
except by anthropic considerations~\cite{GLV04}).  What it does seem to 
call for, rather conclusively, is the abandonning of the more restrictive 
condition $P>0$ that is  usually postulated as a necessity for avoiding 
microscopic instability in a perfect fluid model. 
 
\section{The need for a solid model}

In a perfect fluid model that is barotropic, meaning that its pressure
$P$ is a function just of its mass-energy density $\rho$, what is actually 
needed for avoiding microscopic instability is that there should be a real 
value for the sound speed $c_{\rm _S}$ as given (even in a relativistic 
model) by Newton's well known formula
\be c_{\rm _S}^{\, 2}=\frac{{\rm d} P}{{\rm d}\rho}\, .\label{2}\fe
In the particular case of a relativistic polytrope, meaning a model for 
which the density is proportional to a power of a conserved number density 
$n$ say, i.e. $\rho\propto n^\gamma$ where $\gamma$ is the polytropic 
index (whose value in the familar example of electromagnetic black body 
radiation is $\gamma=4/3$) we shall have
\be P=w\rho\, ,\hskip 1 cm  c_{\rm _S}^{\, 2}=w\, ,\label{3}\fe
in terms of the so called ``equation-of-state parameter'' 
$w$ as defined in terms of the polytropic index by $w=\gamma-1$.
A negative value of ${\rm d} P/{\rm d} \rho$ would imply an imaginary 
value of $ c_{\rm _S}$, which would be interpretable as meaning that 
short wavelength perturbations would undergo rapid exponential growth. 
The condition for avoidance of such instability is evidently expressible 
as
\be \frac{{\rm d} P}{{\rm d}\rho} >0\ \ \Rightarrow\ \  w>0
\, ,\label{5}\fe
which in conjunction with the energy condition (\ref{1}) is equivalent
to the usual supposition $P>0$.

Since the condition (\ref{5}) seems~\cite{CMMS04} to be inconsistent with 
the available cosmological evidence, the conclusion to be drawn is that we 
need a model of a kind more general than a polytropic fluid. Whereas it 
is hard to see how to construct a viable theoretical model that violates 
the energy dominance postulate (\ref{1}), it was pointed out by 
Bucher and Spergel~\cite{BS98} that it is easy to construct a continuum
model that gets round the restriction (\ref{5}) if, instead of assuming 
that it behaves as a fluid, one supposes that it will behave as a solid 
with a sufficiently large rigidity modulus $\mu$.  Generalising results
that are well known in non-relativistic elasticity theory, it was shown
many years ago by the present author~\cite{C73} that, in a relativistic
elastic solid, the speed, $c_{_\Vert}$ say, of longitudinally polarised
propagation modes will be given in terms of the value that it would have 
according to (\ref{2}) in the absence of rigidity by
\be c_{_\Vert}^{\, 2}= c_{\rm _S}^{\, 2}+
\frac {4}{3}\, c_{_\perp}^{\, 2}\, ,\label{6}\fe
where $c_{_\perp}$ is the speed of transversly polarised (shake type) 
modes, which will be given by 
\be c_{_\perp}^{\, 2}=\frac {\mu}{\rho+P}\, .\label{7}\fe
It is evident that these speeds will both be real, and hence that the 
isotropic solid state will be locally stable, provided  not only that the 
rigidity is positive, $\mu>0$, but that it also satisfies 
\be \frac{\mu}{\rho}> -\gamma\, w\, ,\label{8}\fe
which is evidently a more restrictive requirement if $w<0$. It can be seen  
that this stability criterion  will be satisfied for all (negative as well 
as positive) values of $w$ if  
\be \ \frac{\mu}{\rho}>\frac{1}{4}\, .\label{9}\fe

\section{Membranes versus strings}

Having presented the case in favour of a solid model, Bucher and Spergel 
went on to suggest~\cite{BS98,BBS99} that a medium of this kind might 
arise naturally as a large scale average representation of a distribution 
of approximately static cosmic strings of the simple Nambu Goto type 
with tension $T$ equal to energy density, for which the effect of 3 
dimensional averaging would give an effective average tension, meaning a 
negative pressure, that would be one third of the average density, i.e.
\be w=-1/3 \ \ \Leftrightarrow \ \gamma=2/3\, .\label{10}\fe
They also effectively resuscitated an earlier idea (first put forward 
in a very different astrophysical context when the cosmological evidence 
for dark energy dominance was not so strong) of Kubotani and 
collabotators~\cite{Kubotani92,DYK00} by pointing out that from
the point of view of the cosmological evidence a more satisfactory 
agreement would be provided by a distribution of Dirac type membranes, 
for which the tension would also be equal to the energy but for which 
averaging over space directions would give a mean tension twice as large, 
so that one would have
\be w=-2/3 \ \ \Leftrightarrow \ \gamma=1/3\, .\label{11}\fe

As when a gas of particles is treated as a fluid, the continuum description 
of such a distribution of strings or membranes as a coherent medium 
would presumably be valid only at a macroscopic level sufficiently
large compared with some relevant interaction lengthscale.
The question of whether the elastic deformation energy of such an 
isotropic string or membrane distribution really would be sufficient to 
satisfy the stability condition (\ref{8}) was not checked until very 
recently, but the result~\cite{BCCM05} turns out to be positive, at least 
for systems of even type, meaning those for which the branching number at
junctions is even, so that the intersections can be described in terms
of crossing without deflection -- with the implication that local 
equilibrium of a distribution of straight strings or plane membranes
can be preserved by linear deformation. It has been found for such
cases that the effective rigidity modulus describing the large scale 
averaged value of such a distribution will be the same for the membrane 
case as for the string case, having a value given by
\be \frac{\mu}{\rho}=\frac{4}{15}\, .\label{12}\fe
This is clearly sufficient -- though not by a very large margin -- to 
satisfy the general stability condition (\ref{9}), a consideration that 
is important in view of the likelihood that -- even if it is dominated by 
membranes -- a naturally ocurring defect distribution would be likely to 
include a certain proportion of strings (in particlular those forming the 
junctions between membranes) and therefore to be characterised by an 
effective index in the intermediate range $-2/3<w<-1/3$, $1/3<\gamma<2/3$.

The traditional simplified Friedman model of a homogeneous isotropic
universe is governed by an evolution equation expressible (in Planck units 
with ${\rm c}={\rm G}=\hbar=1$) in the well known form
\be H^2=\frac{4\pi}{3}\rho\, ,\label{13}\fe
where $H$ is the Hubble expansion rate as given with respect to the proper 
time $t$ in terms any comoving lenghtscale $\lambda$ by $H=\lambda^{-1}
{\rm d}\lambda/{\rm d}t$. The preceeding considerations suggest that in 
addition to the usual cold matter contribution proportional to a conserved 
number density 
\be n=\lambda^{-3}\, ,\label{14}\fe
say,  and the (in the early stages dominant) black body radiation 
contribution proportional to $n^{4/3}= \lambda^{-4}$, we should add in an 
allowance for a cosmic string contribution proportional to $n^{2/3}=
\lambda^{-2}$ and a cosmic membrane contribution to $n^{1/3}=
\lambda^{-1}$. (In a brane world cosmological scenario one would also 
need extra terms~\cite{BDL00,CU01} to allow for effects of higher 
dimensions that could have been important at very early stages 
in the evolution, but this should not be necessary for the present purpose 
which is to consider the less speculative question of the more recent
evolution, starting not too much before the observationally accessible 
era at which primordial element formation took place).

Under these assumptions the equation of state for the total mass density 
will be expressible in the convenient  form
\be \rho=\frac{3}{4\pi}\left(n^{4/3}+a\, n+ b\, n^{2/3}+c\, n^{1/3}
\right) \label{15}\fe
in terms of just three constant coefficients, $a$, $b$, $c$, of which the
first drops out in the corresponding expression for the pressure, namely
\be P=\frac{1}{4\pi}\left(n^{4/3}-b\, n^{2/3}-2c\, n^{1/3}
\right) \, . \label{16}\fe
The calibration adopted here allows the Hubble equation to be written in 
the simple explicit form
\be \frac{{\rm d} t}{{\rm d}\lambda}=\frac{\lambda} {\sqrt{1+
a\,\lambda+b\,\lambda^2+c\,\lambda^3}}\, .\label{17}\fe
These expressions implicitly fix the normalisation of the comoving 
lengthscale $\lambda$, which  can be seen to be interpretable as a mean 
black body radiation wavelength, i.e $\lambda\approx 2\pi\Theta^{-1}$ where 
$\Theta$ is the temperature, whose present day value, $\Theta_{\rm c}$ say, 
corresponds to a wavelength in the millimeter range, which in Planck 
units means something like $\Theta_{\rm c} \approx10^{-31}$. The 
parameter $a$ therefore represents the temperature at which cold matter 
and radiation densities were comparable, which appears to be somewhat 
less than the Rydberg recombination energy, (the exact value depending on 
how many neutrino species are supposed to be present) so that in terms 
of the fine structure constant $e^2\approx 1/137$ and the electron mass 
$m_{\rm_e}\approx 10^{-22}$ it will be given very roughly by 
$a\approx 10^{-1}e^4\, m_{\rm_e}\approx 10^{-27}$. 

The novelty in the model considered here is the introduction of the other
two coefficients $b$ and $c$ of which the latter, namely the membrane 
density coefficient $c$, is presumably the most important. The string 
density coefficient $b$ is likely to be relatively unimportant, because it 
is expected that the string contribution would have been dominated by the 
cold matter and radiation contributions at earlier stages, while it will 
evidently be dominated by the membrane contribution at later stages, and 
already -- if observational appearances are to be believed -- even at the 
present epoch. In order for the membrane term to be comparable with the 
cold matter term as from about now, it is clearly necessary that its 
coefficient should be given in rough order of magnitude by $c\approx 
a\,\Theta_{\rm c}^{\,2} \approx 10^{-89}$.
 
\section{Membrane formation by the Kibble mechanism}

Up to this stage, the line of logic presented  here has followed the
principles originally proposed by Bucher and Spergel~\cite{BS98,BBS99},  
but -- in so far as the mechanism governing the creation of the membranes, 
and the implications for the value of the membrane density coefficient $c$ 
as a function of the membrane tension, $T$, is concerned -- I now have to 
express a somewhat  dissident opinion.  I agree that it is reasonable to 
assume that, after the cosmological temperature $\Theta$ had fallen below 
the value, $\eta$ say, characterising the symmetry breaking phase 
transition by which they were formed, the membranes would have rapidly 
settled down ``to an equilibrium distribution which is swept along by the 
Hubble flow''. However I do not agree with the supposition, expressed in 
the same sentence~\cite{BBS99}, that there would initially have been ``one 
wall per horizon volume'', i.e. the minimum that is causally conceivable.
The reasonning below would have it that the correct value should be not one 
but more like $10^{10}$.

A supposition to the effect that there is initially a single defect per 
horizon volume is of course the usual starting point for 
studies~\cite{Shellard95} of free oscillations, radiation damping, et 
cetera, but only after, not before, the Kibble transition at which the 
assumption that the distribution is frozen into the Hubble flow ceases to 
be valid. For defects formed when $\Theta\approx\eta$  it is 
expected~\cite{VS94} that the Kibble de-freezing transition will occur when 
$\Theta\approx \eta^2$, which will happen at a fairly early stage for very 
heavy defects such as GUT strings (as characterised by $\eta\approx 
10^{-3}$).  However for the much lower values of $\eta$ that are relevant 
in the present context, the temperature will still be much too high, even 
at the present epoch, for the Kibble transition to have occured at all. 
This simplifies the analysis in many ways -- justifying the neglect of 
computationally awkward effects such as gravitational radiation reaction, 
and allowing us to assume with confidence, at least as a first 
approximation, that the distribution will indeed be ``swept allong by the 
Hibble flow'' --  but it also means that it is quite inappropriate to assume 
that there was initially  just ``one wall per horizon volume''.

The correct supposition, or at any rate what is usually assumed in
discussions~\cite{VS94} of Kibble's defect formation mechanism, is that 
there would indeed have initially been one defect per correlation volume, 
but that this volume would have been the cube of a correlation length, 
$\xi$ say, that would initially have been much smaller the horizon length 
scale $H^{-1}$  (a causal limit value that it can attain only at a much 
later stage, if and when the Kibble transition temperature is reached). The 
usual Kibble ansatz~\cite{CBDS00} (based on random walk considerations) for 
the initial value of the relevant correlation length is that it would have 
been the geometric mean of the horizon scale $H^{-1}$ and the thermal 
lengthscale, $\lambda\approx\Theta^{-1}$ at the epoch of the phase 
transition, which gives
\be \xi\approx \eta^{-3/2}\, ,\label{18}\fe
(whereas the ansatz used by Bucher and Spergel amounted to taking the 
much larger initial value $\xi\approx \eta^{-2}$). 

In the simple cosmic string case the subsequent increase in the correlation
scale (which eventually catches up with the Hubble radius at the Kibble 
transition) will entail erasure of structure on smaller scales (small 
wiggles will be damped out and small loops will contract to nothing) 
but in the kind of membrane network envisaged here the damping process 
characterised by the expanding lengthscale $\xi$ will not entirely destroy 
the structure on smaller scales but merely freeze it into a configuration 
with locally minimised energy, which one would expect to resemble a 
crystaline or glass-like lattice of cellular vacuum domains characterised 
by a  ``rigging'' lengthscale $\ell$ say. This rigging lengthscale would 
initially be the same as the damping lengthscale $\xi$ but it would expand 
more slowly, merely being dragged along by the Hubble flow and therefore 
given at later times by
\be \ell\approx \eta^{-1/2}\Theta^{-1}\, .\label{19}\fe
(This is to be compared with the corresponding formula for the evolution 
of the damping scale, which according to the usual analysis~\cite{VS94} 
will be given by $\xi\approx\eta\,\Theta^{-5/2}$.)

Since the surface density is given by the membrane tension $T$, the
mass energy associated with a single wall of a single cellular vacuum 
domain will given in order of magnitude by $T\,\ell^2$. This leads to 
an expression of the form
\be \rho\approx \frac{T}{\ell}\, .\label{20} \fe
for contribution of the membrane distribution to the cosmic mass density, 
which according to (\ref{19}) will therefore be given as a linear function 
of the cosmic temperature by the order of magnitude estimate
\be \rho\approx \eta^{1/2} T\,\Theta\, ,\label{21}\fe
which is interpretable as meaning that the coefficient $c$ in (\ref{15}) 
will be given roughly by
\be c\approx \eta^{1/2} T\,  .\label{22}\fe
(It is to be remarked that this differs from the corresponding estimate 
in the analysis of Bucher and Spergel~\cite{BS98,BBS99} by the inclusion 
of the factor $\eta^{1/2}$, which can be expected, as will be shown below, 
to have a magnitude comparable with  $10^{-10}$).

\section{Estimation of the membrane tension}

 The kind of model envisaged in the foregoing analysis is illustrated
by a a modified O(N)  model of the type considered (albeit in a different
astrophysical context involving symmetry breaking at much higher energy 
scales) by Kubotani and collaborators~\cite{Kubotani92}, of which the 
simplest relevant case  consists of an ordinary O(3) model for a set of real 
scalar fields  $\Phi_i\, ,$ $i=1,2,3$,  with an explicit symmetry breaking 
term proportional to a small parameter $\varepsilon$, as given by a 
potential energy density term given by an expression of the form
\be V\ \propto\  (\Sigma_i\Phi_i^{\,2}-\eta^2)^2+
\varepsilon\Sigma_i\Phi_i^{\, 4}\, .\label{23}\fe
If $\varepsilon< 0$ this potential has 6 minima, corresponding to 6 
distinct degenerate vacuum states given for different choices of index 
$j=1,2,3$ by  $\Phi_j=\pm\eta/u$  and by $\Phi_i=0\, , i\neq j$, where 
$u$ is an order of unity factor given by $u^2=1+\varepsilon=
1-\vert\varepsilon\vert$. On the other hand if $\varepsilon>0$ there will 
be 8 minima given by taking all the fields to have the same amplitude, 
$\vert \Phi_i\vert=\eta/u$ but with different combinations of positive 
or negative sign, in terms of an order of unity factor that will be
given in this case by $u^2=3+\varepsilon$.

Ordinary 3-dimensional space, with Cartesian coordinates $x^i$, will have a
natural periodic map, given by $u\Phi_i=\eta\,{\rm sin}\{\pi\, x^i/2\ell\}$,
taking it onto a cube in the configuration space with the calibration 
adjusted so that the cube sides extend just as far as these minima. This 
gives a field configuration in which the vacua will arrange themselves as 
a face centered cubic lattice in the 6-fold case, i.e. when $\varepsilon<0$, 
and as an ordinary body centered cubic lattice in the 8-fold case
(see Figure 1)  i.e. when $\varepsilon>0$. For a given mesh scale $\ell$, 
the energy will be minimisable by a continuous adjustment whereby  the 
vacuum region expands to fill nearly all the interior of each cell of the 
lattice, leaving only a thin boundary layer characterised by a membrane 
thickness, $\delta$ say, given in order of magnitude by
\be \delta\approx \vert\varepsilon\vert^{-1/2}\eta^{-1} \, ,\label{24}\fe
so that the corresponding surface energy density and membrane tension  
will be given by
\be T\approx \vert\varepsilon\vert^{1/2}\eta^3\, .\label{25}\fe

In the sixfold case characterised by $\varepsilon<0$
there will be string-like boundaries where 3 membranes meet
(and monopole like junctions where 8 such strings meet) which means
that the system will be classifiable as of ``odd'' type, and hence
that it does not satisfy the conditions needed for the demonstration
of stability referred to above~\cite{BCCM05}. The doubtful stability
of such a system disqualifies it from being a plausible candidate
mechanism for solving the dark energy problem under consideration here 
(though it does not exclude it from relevance in the much higher energy
context of the large scale structure formation considered by
Kubotani and collaborators~\cite{DYK00}). 

\begin{figure}
\label{firstfig}
\centering
\epsfig{figure=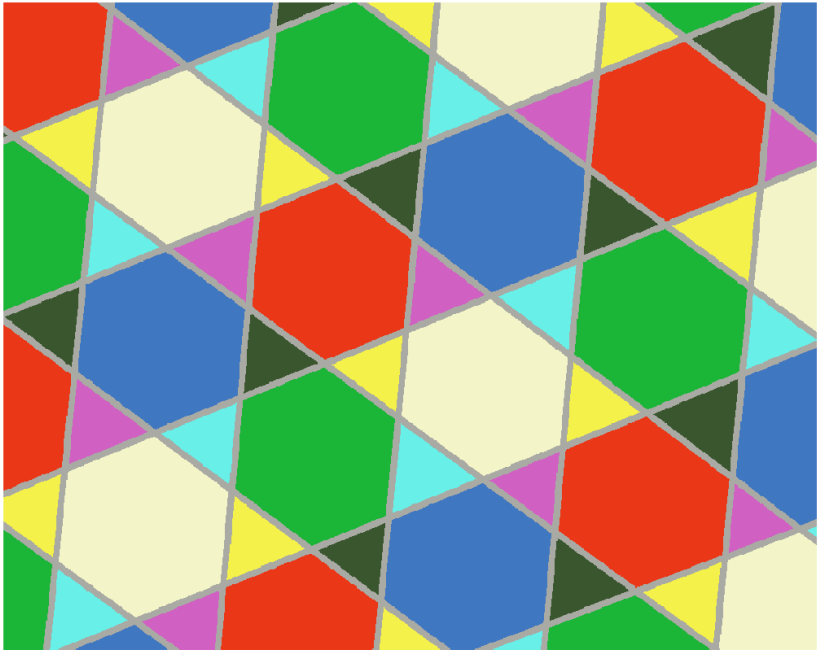, height=9 cm}
\caption{Section through ``8 color'' body centered cubic lattice.}

\end{figure}

For setting up an effectively rigid solid structure, the positive 
alternative $\varepsilon>0$ is more promising, since it provides a  
$2^{\rm N}$ fold  system that will always be of evenly intersecting 
type, not only when N itself is even, but also when N is odd. In 
particular, for N=3, one gets an eightfold system of ``even'' type 
in which the variously ``colored'' vacuum states are given by the 
different choices of sign for the triplet 
$\{\Phi_{_1},\Phi_{_1},\Phi_{_1}\}$, which are listable 
in standard (natural) color notation as a set of 4 opposing (nowhere 
neighbouring) pairs, namely
``black'' $\{-,-,-\}$, and ```white'' $\{+,+,+)\}$, ``red'' $\{+,-,-\}$
 and  ``cyan'' $\{-,+,+\}$,  ``green'' $\{-,+,-\}$ and ``violet'' 
$\{+,-,+\}$, ``blue'' $\{-,-,+\}$ and ``yellow'' $\{+,+,-\}$
(see Figure 1). In this case there will be string-like boundaries where 4 
membranes meet (and monopole like junctions where 6 such strings meet) so 
it is clear that the system will be classifiable as of the ``even'' type to 
which the simple rigidity analysis referred to above~\cite{BCCM05} is 
applicable. Taken together such  system of membranes, strings, and 
monopoles will constitute a dynamically well behaved rigging structure in 
the sense~\cite{Carter95} described in section 2. The latter (string and 
monopole) constituents will become less and less important as the system 
expands, so the rigging structure  will end up by being entirely 
dominated by the domain wall segments, which would be triangular in 
the 6-fold ``odd'' case, but will be square in the 8-fold ``even'' case 
that is relevant.

The lattice structure obtained in this way would be highly isotropic for 
large values of N. Even for such a low value as N=3, although its 
mechanical properties would only be approximately but not exactly 
isotropic at a local level, the rigging system would still provide an 
effectively isotropic stress tensor. It might also be possible to obtain 
mechanically isotropic behaviour on scales with magnitude $\xi$ small 
compared with the Hubble radius $H^{-1}$ but large compared with the mesh 
scale $\ell$ in a disordered system of the kind whose possibility is 
suggested by the 5-fold example described in the next paragraph. 
The observational detectability of the very large scale granulation 
on the scale of such a correlation length has recently been examined by 
Friedland~\cite{Friedland03}, who has thereby derived limits on $\eta$ 
that are remarkably consistent with what has been obtained from the line 
of reasonning described above.

\begin{figure}
\label{secondfig}
\centering
\epsfig{figure=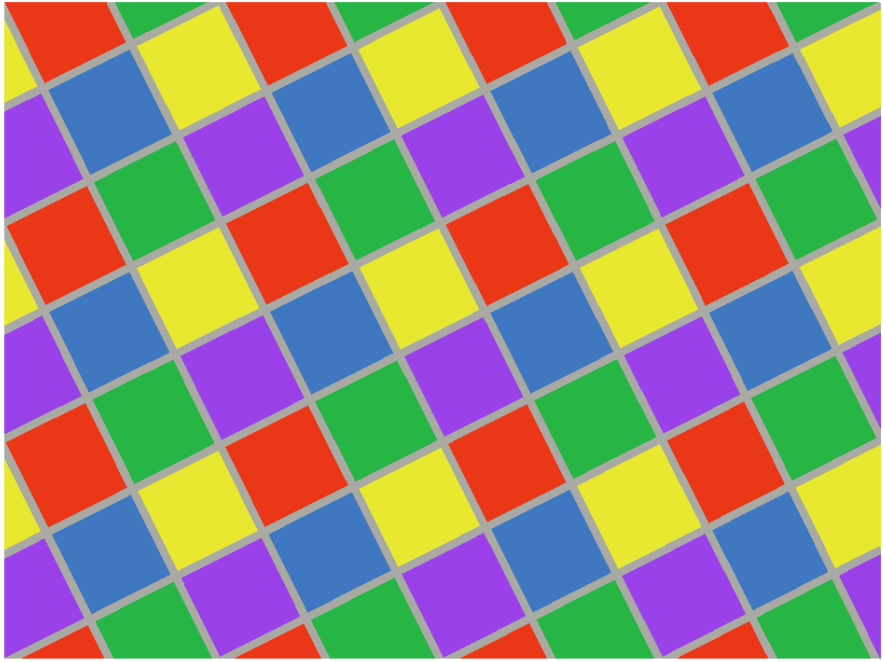, height=9 cm}
\caption{Periodic``5 color'' square tiling lattice with pentahedral symmetry.}

\end{figure}

It is to be remarked that although the simple evenly intersecting systems 
considered above involved an even number of vacuum configurations, it 
is also possible to have systems of ``even'' type having an odd number of
vacuum configurations. A noteworthy example of an ``even'' type system
having 5 distinct vacuum states (related by a discrete pentahedral 
symmetry) is obtainable from an ordinary  U(1)$\times$ U(1) model involving
4 real scalar fields that combine to form a pair of complex fields
\be\Phi_{_1}+i\Phi_{_2}=|\Phi|{\rm e}^{i\phi}\, ,\hskip 1 cm
\Psi_{_1}+i\Psi_{_2}=|\Psi|{\rm e}^{i\psi}\, ,\fe
by adding a symmetry breaking term with, let us say, a positive 
coefficient $\varepsilon<1$, to the usual quartic potential so that it 
acquires the form
\be V \propto (|\Phi|^2-\eta^2)^2+ (|\Psi|^2-\eta^2)^2+
\varepsilon|\Phi|^2|\Psi|^2\Big({\rm cos}(\psi+2\phi)+{\rm cos}(2\psi-\phi)
\Big)\, .\fe
This has minima with $|\Phi|^2=|\Psi|^2=\eta^2/(1-\varepsilon)$ 
at vacuum configurations for which the values of the angle pair 
$\{\phi,\psi\}$ can be listed in terms of arbitrarily ascribed 
``colors'' as the obvious combination $\{\pi,\pi\}$, ``green'' say,
together with  $\pm\{\pi/5, 3\pi/5\}$, ``red'' and ``blue'' say,
 and finally $\pm\{3\pi/5,-\pi/5\}$, ``violet'' and ``yellow'' say.
Unlike the eightfold system discussed above, this fivefold system
is not subject to any restrictions on which pairs of ``colors'' can 
be neighbours to each other, whether diagonally across a junction or 
face to face directly across a wall. Indeed all conceivable neighbouring 
``color'' relationships are exhibited in the pentahedrically symmetric
square ``tiling'' pattern induced on a 2-dimensional plane (see Figure 2) 
by taking the phases $\phi$ and $\psi$ as Cartesian coordinates -- which 
suggests that as well as being able to form a periodic crystaline lattice, 
such a system also offers the possibility of forming a less regular 
lattice structure (see Figure 3)

The question of irregular generalisation of this 5-fold square tiling 
example raises the problem of whether, to color a generically disordered 
2 dimensional tiling geometry, subject to the condition that only quadruple 
(crossroads type) junctions are allowed, and that two tile domains of the 
same ``color'' can never be neighbours, a maximum of 5 colors will always 
be sufficient. This 5 color ``even'' type tiling problem is to be compared 
with the corresponding ``odd'' type tiling problem, for which the condition 
is that only triple junctions are allowed, in which case it is well known, 
though not so easy~\cite{Horgan93} to prove, that 4 colors will suffice -- 
with the corollary that 4 colors would also suffice when the junctions 
are of quadruple type if the rules were relaxed to allow diagonal (though 
still not face to face) contact between tile domains of the same color.

\begin{figure}
\label{thirdfig}
\centering
\epsfig{figure=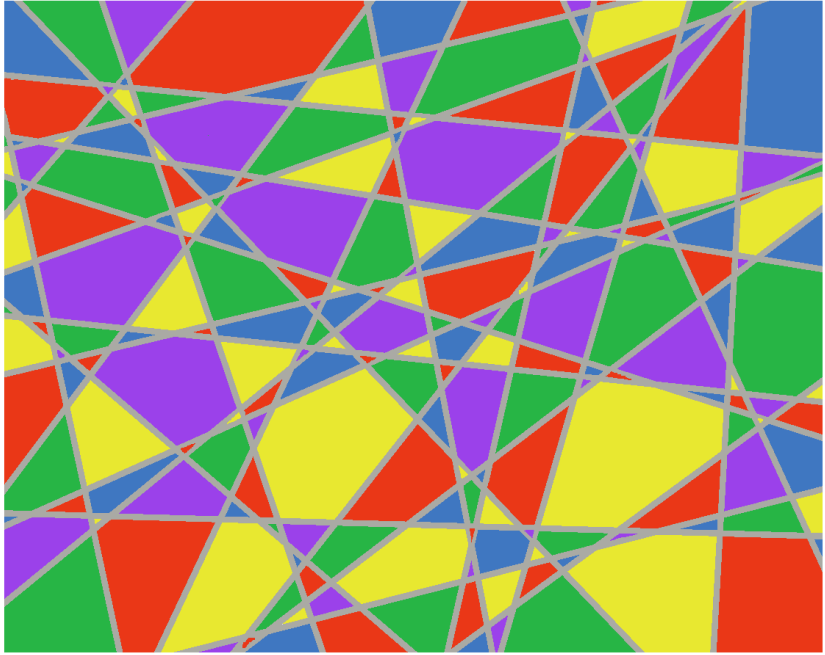, height=9 cm}
\caption{Disordered ``5 color'' tiling lattice.}

\end{figure}

\vfill\eject
                                
\section{Conclusions}

The kind of frozen (locally static) rigging system described in the
preceding section will provide roughly what seems to be 
needed~\cite{CMMS04} to account for the cosmological appearance of dark 
energy provided the microscopic field theory parameters $\eta$ and 
$\varepsilon$ are such that the corresponding value of the coefficient 
$c$ in (\ref{15}), which according to (\ref{22}) and (\ref{24}) will be 
given by 
\be c\approx \vert\varepsilon\vert^{1/2}\eta^{7/2}\, , 
\label{26}\fe
is consistent with the order of magnitude suggested by the observational
considerations described in section 3, namely
\be c\approx 10^{-1}e^4\, m_{\rm_e}\Theta_{\rm c}^{\,2}
\approx 10^{-89}  \, ,\label{27}\fe
in the Planck units that have been used throughout. Due to the high 
power of its involvement in (\ref{26}), the required value for the mass 
scale $\eta$ is not very sensitive to uncertainties in $c$ or in 
$\varepsilon$, and works out to be not so very much smaller that the value 
obtained from the causal limit assumption used originally by Bucher and 
Spergel~\cite{BS98,BBS99}. The mesh-scale $\ell$ as given by (\ref{19}) 
is even less sensitive, coming out to be larger than the black body 
wavelength by a factor of the order of $10^{11}$ --  which is comparable 
with interstellar distances of a few light years at the present epoch -- 
while the mass scale itself will be given by
\be \eta\approx \vert\varepsilon\vert^{-1/7} 10^{-25} \, .\fe
It can be seen that this will be significantly, but not enormously
smaller that the electron mass, $m_{\rm e}\approx 10^{-22}$, assuming 
that $\varepsilon$ has a value that is much -- but not too much -- smaller 
than unity. 

The meaning of this is that the relevant membrane forming phase transition  
would have to have occurred somewhere round about the Kev range, prior to 
the recombination transition whereby the universe became optically 
transparent but subsequently to the cosmological epoch of  primeval element 
formation, which is the earliest stage for which it can be claimed that we 
have detailed observational information. (Its occurrence in such a 
relatively accessible energy regime encourages me to conjecture that the 
phase transformation in question may have something to do with the breaking 
of the approximate isospin symmetry between up and down quarks.)

\bigskip

{\bf Acknowledgements}: I am indebted to Robert Brandenberger, Richard 
Battye, Martin Bucher, Anne Davis and Alex Vilenkin for many instructive 
discussions.

\vfill\eject

\end{document}